# Electromagnetic manipulation of sub-500 Da biomolecules

**Marco Riccardi** and Olivier J.F. Martin

*Laboratory of Nanophotonics and Metrology, Swiss Federal Institute of Technology,*

*CH-1015, Lausanne*

E-mail: marcori94@gmail.com

Abstract

The manipulation of nanoscale matter has the potential to revolutionize a variety of fields across nanoscience and technology. Here, we demonstrate experimentally and characterize numerically a device that combines the benefits of dielectrophoresis (DEP) – long-range and strong trapping forces – with those of plasmonic tweezers – high sensitivities – to achieve a remarkable efficiency in the trapping and sensing of metallic nanoparticles and biomolecules. In particular, we show the DEP trapping and surface enhanced Raman scattering characterization of bovine serum albumin and Rhodamine B, thus extending the applications of tweezing devices to molecules having masses of only a few hundreds of Da. This range covers virtually any molecule relevant for life, from tiny oligopeptides to large proteins. This pushes our manipulation capabilities deep into the realms of efficient single-molecule biosensing and quantum science, providing a powerful platform to probe matter at the nanoscale.

# Introduction

One of the ultimate goals of nanotechnology is the comprehensive manipulation of nanoscale matter, granting full control over the position and movement of a wide range of materials like, for example, nanoparticles, quantum dots and biomolecules. Such a technology has the potential to revolutionize a variety of fields including manufacturing,[1,2] quantum science and technology [3,4] and point-of-care devices,[5,6] and is therefore of the utmost importance for the development of our society. Meeting these expectations requires devices able to generate strong and long-range forces that can trap and sense analytes of different sizes in a wide range of concentrations and chemical conditions.[7,8]

Among the several trapping approaches that have been proposed in the literature to immobilize different kinds of analytes on a substrate,[9–11] those based on electromagnetic forces have received particular attention from the scientific community. These forces stem from the creation of induced multipolar moments in the analyte, allowing its manipulation with non-uniform electric fields.[12–15] The latter can be generated by employing low frequency oscillating fields to bias a pair of microelectrodes, as schematically shown on the left side of Fig. 1(a), or by employing optical radiation to excite plasmonic resonances inside isolated metallic nanostructures, as shown on the right side of Fig. 1(a). Depending on the approach chosen, one talks about *dielectrophoresis* (DEP) [16–18] in the first case and about *plasmonic tweezers* [19–21] in the second. Both approaches have found similar applications for the trapping of different analytes such as particles[22–25] and large molecules.[26–29]

Plasmonic tweezers, in particular, have attracted significant interest thanks to their ability to act as highly sensitive sensors,[30–32] making them an appealing technology for point-of-care devices. However, trapping with plasmonic tweezers suffers from a significant limitation, which is the very localized nature of the field inhomogeneities around the structures. Indeed, as shown in the Supporting Information – Figure S1 – the optical field gradient generated by a plasmonic tweezer does not reach as far into the surrounding medium as the DEP one. As a consequence, such systems can only manipulate analytes that are in close proximity to the trapping hotspot and their use is therefore restricted to highly concentrated samples – 1-0.1 μm$^{-3}$ for particles[24] and mM for molecules[33] – or to applications where a highly efficient on-demand trapping is not required. Different strategies have been proposed to overcome this issue,[34] which all rely on complex protocols for the fabrication of optofluidic devices[35,36] or on the exploitation of thermal effects, which are particularly sensitive to the sample's properties and are therefore hard to control.[37] As a consequence, the facile realization of a reliable plasmonic tweezer system for efficient analyte trapping has remained elusive, although the intrinsically small size of plasmonic systems makes them the ideal platform to probe matter at the nanoscale, despite their limited trapping range. On the other hand, as we discuss later in this contribution, the use of electrodes to generate the DEP field easily overcomes this limitation and allows the creation of a non-uniform electric field further away from the electrodes, which produces long-range trapping forces able to manipulate analytes down to pM concentrations. Unfortunately, the use of microelectrodes generally restricts DEP to the manipulation of microscale analytes, although advances in nanoelectrodes fabrication techniques have recently allowed the control of sub-50 nm particles and molecules – having masses in the kDa and MDa ranges[38–41] – sometimes with integrated sensing capabilities.[42–44]

To give a unified perspective of both low and high frequency molecular trapping platforms, Fig. 1(b) lists, with data from previous works available in the literature,[39–43,47–59] the concentration and mass of molecules that have been manipulated so far with DEP and plasmonic tweezers. This figure is very telling as it shows that, while one would ideally perform detection of very small molecules in highly diluted samples, detection of small masses can only be achieved at the expenses of increasing the molecular concentration. Vice versa, very scarce amounts of analyte can be detected as long as its mass is big enough to generate a strong dipole moment that allows for electromagnetic manipulation. Interestingly, the red circle encloses those works that employ optical fields for both the trapping and detection of the analyte,[60] which can clearly only occur at high molecular concentrations.

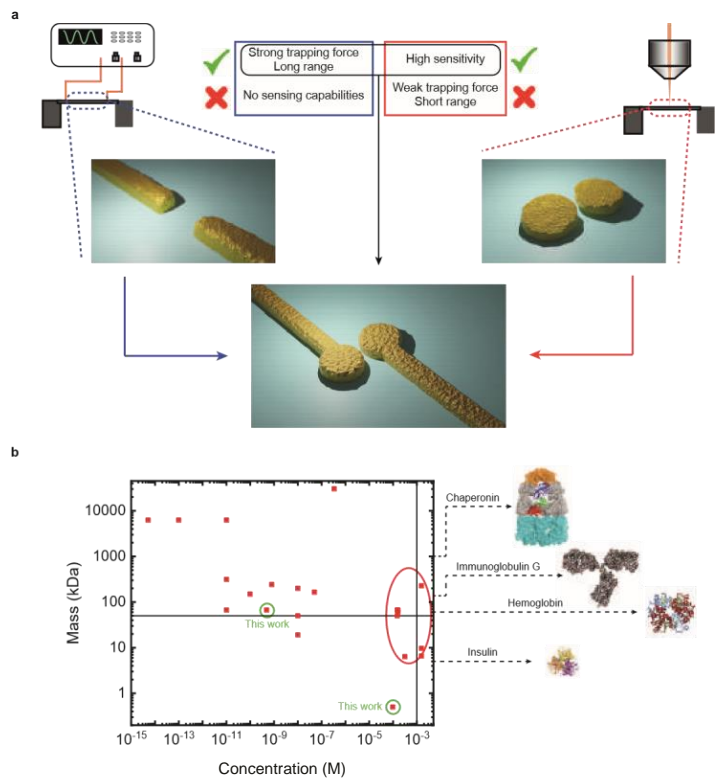

**Figure 1:** (a) DEP, on the left, employs electrodes to generate strong and long-range trapping forces, but completely lacks any sensing capabilities. On the other hand plasmonic tweezers, on the right, provide weak trapping forces but high sensitivities. By combining the two, it is possible to realize a set of plasmonic electrodes able to manipulate both the low and high frequency fields needed, respectively, for DEP trapping and plasmonic sensing. (b) Mass vs. the concentration of manipulated molecules in the literature, with the mass of some biologically relevant proteins shown on the right for reference. The two black lines represent the average size and concentration of proteins in the human body,[45,46] while the red circle encloses works that employ purely optical fields for both the trapping and detection of molecules. Finally, the green circles indicate the work presented in this paper.

Sensing smaller amounts of analytes requires the integration, in the optical setup, of different delivery systems that can increase the trapping yield, such as the use hydrodynamic forces or low frequency electric fields. Here, we tackle this challenge by employing a gold disk dimer as a plasmonic tweezer biosensor. To boost its trapping efficiency, we connect each disk to a long and thin nanorod that is electrically biased to produce a DEP trapping field.[61,62] As shown in Fig. 1(a), this allows the creation of a pair of plasmonic electrodes that are able to manipulate both low and high frequency electric fields, to be used respectively for the DEP trapping and optical sensing of nanoscale analytes. Notably, we employ surface-enhanced Raman spectroscopy (SERS) to report the efficient DEP manipulation of bovine serum albumin (BSA, 66 kDa) at concentrations as low as 500 pM and the DEP trapping of Rhodamine B (479 Da) which, to the best of our knowledge, has never before been reported in the literature. This pushes the capabilities of electromagnetic manipulation to a new length scale, where analytes are composed of less than one hundred atoms, paving the way to the realization of ultra-sensitive biosensors that have the potential to revolutionize the fields of molecular detection and manipulation. To give some perspective, we point out that by considering the average mass of an amino acid ($\simeq$ 130 Da), the manipulation of rhodamine implies the ability to control any oligopeptide composed of just four amino acids, virtually comprising any of the protein found in eukariotic, bacterial and archaeal species.[63] Furthermore, the manipulation of BSA demonstrates how our device is also able to manipulate molecules across several orders of magnitude of mass, making these plasmonic electrodes a powerful tool for the biomedical and bionalaytical sciences. In particular we stress the fact that, compared to other aforementioned works concerned with the integration of DEP into an optical sensing platforms, our approach is the first to employ plasmonic structures as working DEP electrodes, instead of more standard microelectrodes with nanoscale protrusions. Hence, we can exploit the enhanced plasmonic interaction between our electrodes and light to provide a *sensitive* and *tunable* detection of different analytes. Recently, a similar approach has also been proposed employing nanoscale apertures inside a gold film.[59] However, the DEP electrodes employed in this work are placed several hundreds of microns apart, resulting in a trapping force several orders of magnitude weaker than the one documented here. This, in turn, allows us to manipulate molecules about 10 times smaller than what reported by these authors, as we will discuss later.

## Theoretical analysis

We start our treatment by characterizing the ability of these plasmonic electrodes to generate both low and high frequency inhomogeneous electric fields for the trapping of gold nanoparticles with a diameter of 40 nm. Given their small size, only a dipolar moment is induced in these particles, which gives rise to a time-averaged force equal to[15,64,65]

$$\mathbf{F} = \frac{\alpha}{4} \nabla |\mathbf{E}|^2 \qquad (1)$$

where α is the real part of the particle polarizability and $|\mathbf{E}|$ is the amplitude of the electric field. This field can be any low or high frequency field, and the theory outlined in this section applies to both DEP and optical trapping.[15] Equation (1) shows that it is possible to use an inhomogeneous electric field to manipulate a dipolar particle, which moves in space towards the regions of stronger field provided that α > 0, i.e. when the particle is more polarizable than the surrounding medium. Moreover, the force on the analyte disappears for $\nabla |\mathbf{E}|^2 = 0$ and the particle can be trapped at a location where a field maximum exists. For a spherical particle of radius *r*, we can write the polarizability α as[66]

$$\alpha = \Re\left[\frac{\alpha_{CM}}{1-\frac{ik^3\alpha_{CM}}{6\pi\varepsilon_m}}\right] \quad (2)$$

where $i$ is the imaginary unit, $k = 2\pi/\lambda$ with $\lambda$ the field wavelength, and $\varepsilon_m$ the permittivity of the medium surrounding the particle. Finally, $\alpha_{CM}$ is the electrostatic Clausius-Mossotti polarizability[64]

$$\alpha_{CM} = 4\pi r^3 \varepsilon_m \Re\left(\frac{\varepsilon_p - \varepsilon_m}{\varepsilon_p + 2\varepsilon_m}\right) \quad (3)$$

with $\varepsilon_p$ ep the particle permittivity. Clearly, at the low field frequencies employed in DEP, $k \to 0$ and, consequently, $\alpha \to \alpha_{CM}$. Therefore, for the case considered here of a 40 nm gold nanoparticle immersed in water ($\varepsilon_m \approx 80\varepsilon_0$, with $\varepsilon_0$ being the vacuum permittivity) $\varepsilon_p \gg \varepsilon_m$ holds and Eq. (3) yields $\alpha_{CM} = 7.12 \cdot 10^{-32} \text{Fm}^2$. On the other hand, at optical frequencies, $\alpha_{CM}$ is calculated using refractive index data from the literature,[67] and Eq. (2) yields $\alpha = 1.80 \cdot 10^{-33} \text{Fm}^2$ at 308 THz, corresponding to $\lambda = 974$ nm. This readily demonstrates one of the benefits of employing low frequency fields to trap gold nanoparticles in a water solution, which is that their low frequency polarizability is more than one order of magnitude larger than that at optical frequencies and therefore generates a stronger force. On top of this, we have already mentioned that DEP is in principle also able to generate an inhomogeneous field $\mathbf{E}_{DEP}$ further away from the electrodes. This difference stems from the different generation mechanisms for the DEP and the optical fields: while electrodes are used to produce a low frequency field, propagating electromagnetic waves are employed to excite plasmonic modes in the tweezer. As a consequence the presence of the inhomogeneous optical near field, responsible for the generation of the trapping force, is quickly concealed by the background excitation field, which has a constant intensity, and therefore prevents trapping far away from the electrodes. This can be better appreciated in Fig. 2(a), which shows the intensity of the low and high frequency fields above the electrodes. Here and throughout the paper, the fields are computed with finite elements using COMSOL Multiphysics 5.6, where low frequency fields are created by biasing the electrodes with a sinusoidal voltage having an amplitude of 7 V and a frequency of 1 MHz, while high frequency fields are computed for plane wave illumination coming from the $-z$ direction and polarized along $x$, having a wavelength of 974 nm and a power density of 3 mW/µm². We can see in this figure that, for example for $z = 1$ µm, $\partial_z|\mathbf{E}_{DEP}|^2 \neq 0$ while $\partial_z|\mathbf{E}_{OPT}|^2 = 0$; with $\mathbf{E}_{DEP(OPT)}$ being the DEP (optical) field interacting with the plasmonic tweezer and $\partial_z$ the partial derivative along the $z$-direction. Clearly, under such conditions, the optical force exerted on the analyte at $z = 1$ µm will be zero, while DEP fields will still be able to manipulate the particle.

The advantages of employing low frequency fields are further evidenced by the calculation of the trapping potential 5 nm above the disks, along the dimer axis, as shown in Fig. 2(b). For the case analyzed here, the DEP trapping well is about four orders of magnitude deeper than the optical one (note the different vertical axes), which barely reaches twice the value of the Brownian energy $k_B T$ at room temperature, with $k_B$ being the Boltzmann constant and $T$ the temperature of the system. This indicates that, while trapping of gold nanoparticles with plasmonic tweezers is possible, the trapped particle can still easily escape from the tweezer. On the other hand, once the same particle is trapped with low frequency fields, it is stably anchored by the strong DEP force that provides therefore a more reliable trapping mechanism. This is manifested also by the higher stiffness of the DEP trap – 733 pN/nm – when compared to that of the optical one – 73.5 fN/nm – as calculated from the plot of the trapping force provided in Fig. 2(c). This figure provides additional insights onto the magnitude of the low and high frequency trapping forces: the former takes values up to 20 nN, while the latter barely reaches 2 pN. Altogether, it is therefore clear that the further-reaching DEP field, combined with the higher particle polarizability at low frequencies, results in a DEP trapping force that is stronger and has a longer range than its high frequency counterpart. This is illustrated in Fig. 2(d) where the trapping volume – defined as the volume of space where the trapping force overcomes the Brownian force $F_B = k_B T/2r = 103.5$ fN [68,69] – is provided for both the DEP and optical forces. It is obvious that a plasmonic tweezer can only exert a force strong enough to manipulate a particle in the very close proximity of the structures, while DEP is able to affect the dynamics of particles as far as about 550 nm from the dimer. This results in a trapping volume of 0.348 µm$^3$, as opposed to 0.003 µm$^3$ for the optical force.

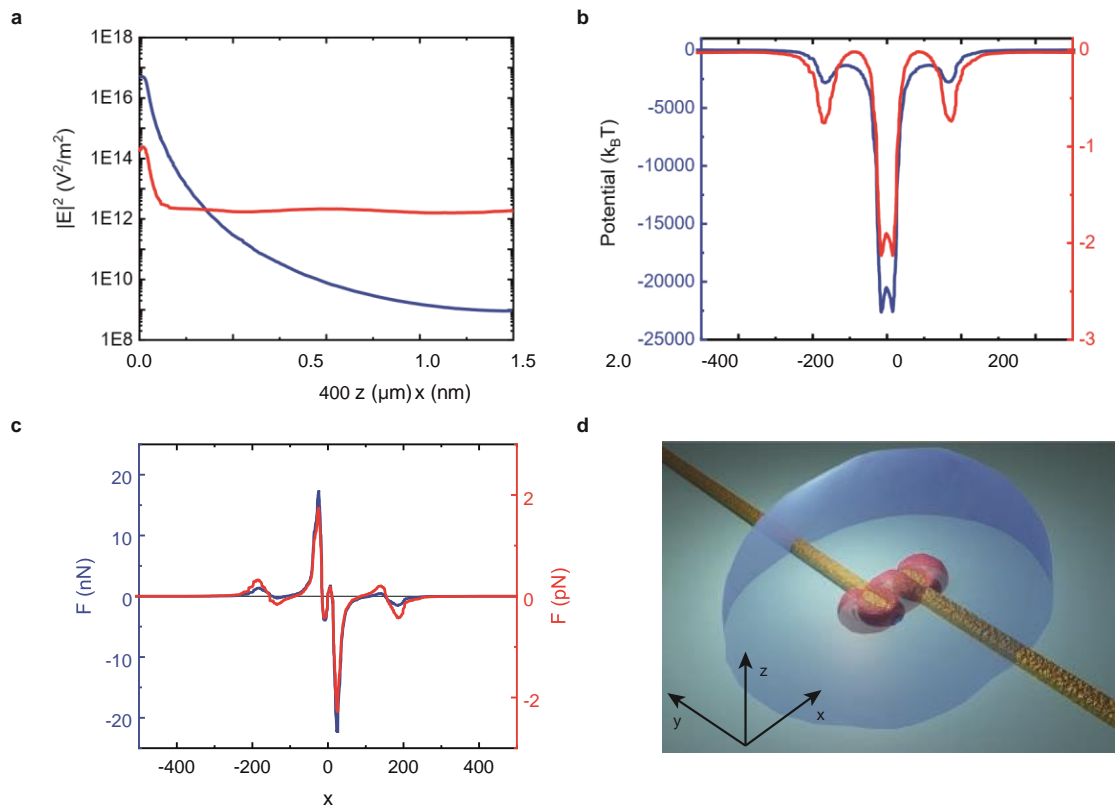

**Figure 2:** Comparison between low (blue) and high (red) frequency trapping forces. Calculated (a) field decays, (b) trapping potentials and (c) trapping forces. Note the different vertical scales in panels (b) and (c). (d) DEP and optical trapping volumes.

## Experimental results

Let us now confirm experimentally the superiority of DEP over plasmonic trapping by manipulating 40 nm gold nanoparticles diluted in water to a concentration of 0.018 µm$^{-3}$. To study the particle trapping, the plasmonic resonance of the disks is simultaneously probed with white light from a halogen lamp polarized along the dimer axis. This ensures a clear optical signature from the electrodes without generating any high-frequency forces that might affect the trapping dynamics. To this end, the left panel of Fig. 3 shows the temporal evolution of the spectral position of the resonance. We clearly see that, as soon as the DEP trapping field is turned on (vertical solid lines in the figure), the plasmonic resonance readily shifts to longer wavelengths, indicating that a nanoparticle has been trapped. Conversely, when the trapping field is turned off (vertical dashed lines), the particle is released and the resonance returns to its initial spectral position. This demonstrates the ability to perform systematic DEP nanoparticles manipulations, even at these low concentrations, which are more than one order of magnitude lower than those usually reported in the literature for a plasmonic tweezer.[24] One expected consequence of this low concentration was the inability to observe multiple particle trapping events, even when keeping the DEP field on for 20 minutes. Overall, this enhanced trapping efficiency was observed to be a robust property of our system, with a reproducibility greater than 95% across multiple devices.

Accordingly, when DEP is not employed and the disk dimer is simply excited with optical radiation at a power density of 3 mW/µm$^2$, we clearly see on the right side of Fig. 3 that no redshift is induced in the plasmonic resonance. It is evident that the weaker and more confined optical field cannot attract the particles toward the tweezer and is therefore not able to stably immobilize a nanoscale object. This clearly shows the limitation of plasmonic tweezers, which are restricted to diffusion-limited systems and cannot be efficiently exploited for very diluted samples. This constraint can of course be overcome by increasing the analyte concentration or the laser power density, but this in turn limits their use in fields such as early molecular diagnostics,[70] where it is important to manipulate and detect minute quantities of analyte, and in those applications where thermal effects are a concern either for the stability of the device or the properties of the analyte solution.[71,72]

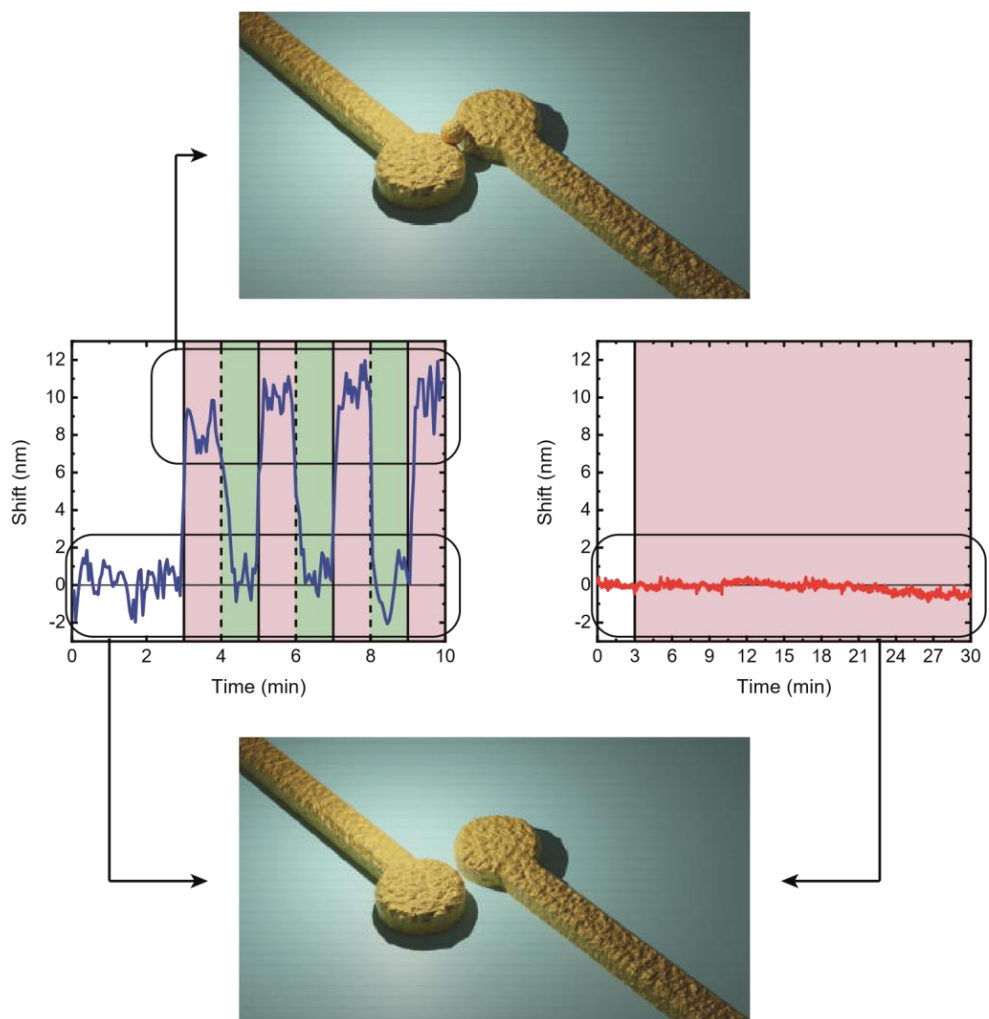

**Figure 3:** DEP (left) and plasmonic (right) trapping of gold nanoparticles. Pink shaded areas indicate the periods when the trapping field is on. Conversely, green shaded areas mark the times when the field is off.

Moreover, for the specific electrode design presented here where the gold structures are anchored to the silica substrate through a molecular adhesion layer – see the Supporting Information – one has to also limit the incident laser power to about 10 mW/µm$^2$ to avoid thermally-induced peeling of the electrodes.[62]

After experimentally confirming the shortcomings of plasmonic trapping and the clear superiority of DEP to perform efficient on-demand nanoparticle control, we now turn our attention to the manipulation of biomolecules, which are more relevant for applications in molecular biology and point-of-care devices.[73] To this end, BSA is a commonly used protein that naturally lends itself to manipulations with electromagnetic fields, thanks to its large mass ($\simeq$ 66 kDa) that ensures the generation of a strong dipole moment. As such, different authors have reported the successful trapping of BSA and its detection exploiting fluorescent labels[74] or refractive index based sensors.[41,75] In this work, after first confirming the ability of low frequency fields in our device to manipulate BSA – see Fig. S3 in the Supporting Information – we then exploit its enhanced interaction with optical radiation in the dimer's gap to record the SERS response of the protein. To this end, the top panel of Fig. 4(a) displays a typical SERS spectrum recorded after concentrating BSA around the dimer with DEP. The spectrum shows a plethora of different peaks, which are easily detectable also thanks to the fact that the laser excitation wavelength employed – 974 nm – is far from the optical absorption band of standard fluorophores. Under this configuration, the SERS signal is therefore not masked by any fluorescence background and we are able to record the full vibrational spectrum of the molecule. For instance, while looking at this spectrum we can identify the stretching S-S bands at 498 cm$^{-1}$ and 550 cm$^{-1}$, the C-H deformation at 682 cm$^{-1}$, the C-S stretching and COO$^-$ deformation at 706 cm$^{-1}$, the N-C$_\alpha$-C stretching at 958 cm$^{-1}$ and the C-C stretching band at 1210 cm$^{-1}$.[76] To further demonstrate the combined DEP manipulation and SERS detection of BSA, the middle panel of Fig. 4(a) shows the evolution of the intensity of the 550 cm$^{-1}$ band before and after the activation of DEP. This figure clearly demonstrates a correlation between DEP and the SERS intensity, which readily increases upon turning on the trapping field as more and more molecules are delivered and concentrated in the SERS detection hotspot, i.e. around the plasmonic disks. We point out that, at the low laser power employed here – 300 µW/µm$^2$ – we expect the molecular trapping to be dominated by DEP, rather than optical, forces. Moreover, the spectrum intensity also depends on the concentration of BSA proteins in the sample solution, with more diluted samples producing a weaker SERS response. For the lowest concentration employed – 500 pM – we also demonstrate, in the bottom panel of Fig. 4(a), the release of the trapped molecules once the DEP is turned off. In this case, the intensity of the 550 cm$^{-1}$ band rapidly decreases as the molecules are now able to diffuse outside the SERS detection zone. We consider the fact that the intensity does not fully return to its initial value, before DEP was turned on, as an indication of BSA molecules having adsorbed on the surface of the gold electrodes either through conjugation of their exposed amino groups or due to dispersion forces. A fit of these data with a modified diffusion equation that takes into account the presence of the substrate,[77] as explained in Section 4 of the Supporting Information, yields a diffusion coefficient for BSA of about $6 \cdot 10^{-7}$ m$^2$/s. This value, which is about four orders of magnitude greater than what one would expect from the Einstein relation, is in agreement with the calculated value for diffusing charged analytes, where their mutual electrostatic repulsion results in an abrupt increase of the diffusion coefficient.[78] BSA molecules, which carry a negative surface charge of about 22.42$e$ at a pH of 8.4,[79] are therefore also expected to experience an enhanced diffusion, as readily detected in our system. To this end we point out that, while other authors have reported the DEP manipulation of BSA solutions at concentrations as low as 1 pM,[41,75] their experimental setup only allowed the detection of a response from the molecules about 10 minutes after turning on the DEP. Contrarily, as shown in Fig. 4(a), our system is able to quickly and efficiently concentrate BSA proteins around the detection zone and readily provides a measurable SERS response right after the activation of the trapping field. This superior efficiency is the result of the combination of the high sensitivity offered by SERS and the ability of our tweezer to generate very strong electric field intensity gradients, up to a value of $3.65 \cdot 10^{25}$ V$^2$/m$^3$ (see Fig. S1 in the Supporting Information), which is among the highest ever reported for a DEP device according to a recent review on the subject.[27]

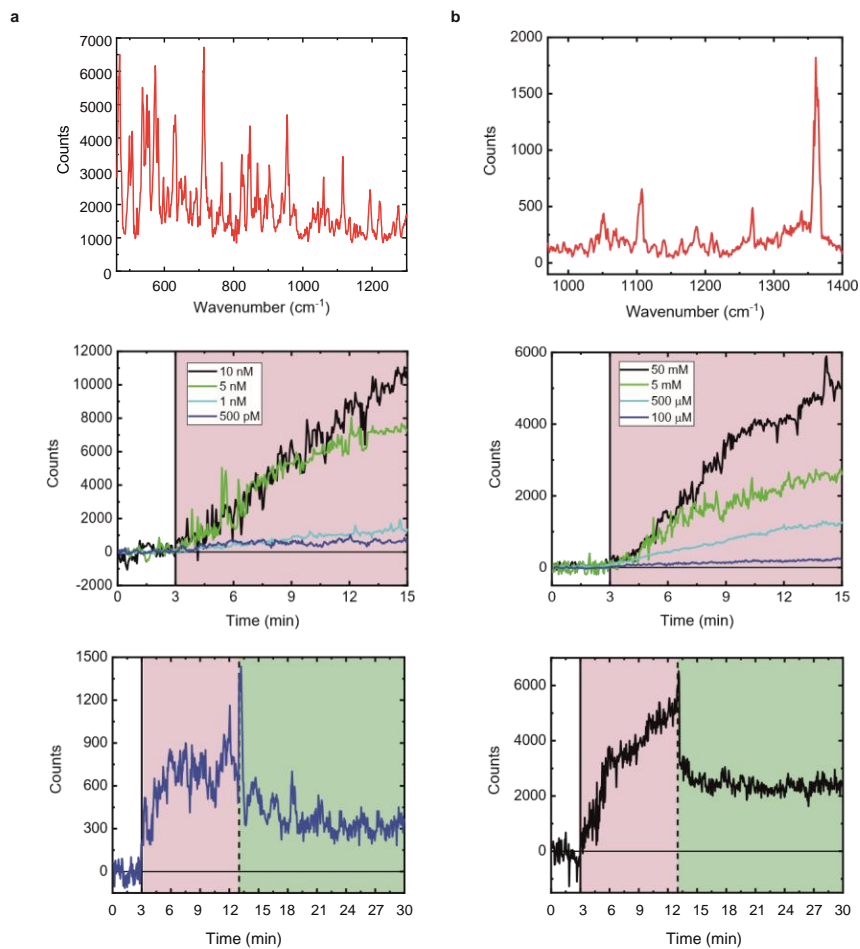

**Figure 4:** DEP trapping and SERS detection of BSA (a) and Rhodamine B (b). On top, a typical molecular SERS spectrum is provided, while the middle panels demonstrate concentration-dependent DEP manipulation. Finally, the bottom graphs show the trapping and subsequent release of the molecules. The pink and green shaded areas represent, as before, the times when the trapping field is, respectively, on or off.

Prompted by these results, we also demonstrate the DEP manipulation of Rhodamine B, which has so far eluded investigations due to its small mass – $\simeq$ 479 Da – that results in the generation of a dipole moment too small to noticeably interact with the electric fields generated in conventional devices. However, by exploiting the superior sensitivity and trapping efficiency of this hybrid tweezer, we are able to provide in the top panel of Fig. 4(b) the measured SERS spectrum of this molecule and confirm its manipulation with DEP. In particular, this figure shows the presence of the C-H, C-C and C-O-C vibrational bands of the xanthene ring of rhodamine around, respectively, 1100 cm$^{-1}$, 1181/1363 cm$^{-1}$ and 1268 cm$^{-1}$.[80–82] By tracking the intensity of the 1100 cm$^{-1}$ band before and after the DEP field is turned on, we are able to demonstrate in the middle panel of Fig. 4(b) the successful DEP trapping of Rhodamine B thanks to the sudden increase of the SERS intensity once DEP is activated. This figure also indicates a dependence on the molecular concentration. The bottom of Fig. 4(b) shows the release of the trapped molecules once the DEP is turned off. In this case, the intensity of the 1100 cm$^{-1}$ band readily decreases as the molecules are now able to diffuse outside the SERS detection zone. As for the case of BSA, we consider the fact that the intensity does not entirely return to its initial value before DEP was turned on, as an indication of rhodamine molecules having adsorbed on the surface of the gold electrodes, for example through electrostatic or dispersion forces.[83] A fit of these data with a modified diffusion equation that takes into account the presence of the substrate[77] yields a diffusion coefficient in the order of $8 \cdot 10^{-10}$ m$^2$/s, in agreement with the value measured through fluorescence correlation spectroscopy.[84]

## Discussion

We have shown how plasmonic electrodes can provide a superior efficiency and sensitivity in the trapping and sensing of metallic nanoparticles and biomolecules. Interestingly, we point out that this enhanced efficiency is somewhat surprising in light of the theoretical modelling of the polarizability put forward at the beginning of this paper. As an example, we computed a DEP trapping volume of 0.348 μm$^3$ for a 40 nm gold nanoparticle which, for the particle concentration of 0.018 μm$^{-3}$ used in the experiments, results in 0.0063 particles present inside the trapping volume – viz. particles that can be manipulated by our electrodes. Despite this extremely low number, Fig. 3 experimentally demonstrates a very efficient trapping dynamics and hints at the fact that the real trapping volume of our electrodes is even larger than the computed one. Indeed, we experimentally show in Fig. S5 of the Supporting Information that the particles trapping volume is at least 50 μm$^3$. This echoes a recent observation made by other authors,[27] which highlighted the fact that DEP trapping of BSA can be performed with weaker electric field intensity gradients than those predicted by the widely-accepted Maxwell-Wigner polarization theory employed here.[15,85] This suggests that the polarizability of BSA is actually larger than what can be calculated with this theory, although no adequate theoretical framework exists yet to account for these observations. It seems that one key to interpret these results is the inclusion in the theoretical treatment of surface effects, and how they affect the analyte induced dipole moment. Such effects include, for example, the presence of charged chemical groups and of an electrical double layer at the analyte's surface, together with the presence of a permanent dipole moment in the analyte,[86,87] all contributing to a strong polarizability increase. Their impact is even more important for nanoscale analytes, where surface phenomena are prominent, while they can be neglected for larger particles whose behaviour is still correctly described by the standard theory. To this end, an empirical theory has been proposed recently to describe the observed DEP trapping of proteins and predicts, for example, an increase of three orders of magnitude for the polarizability of BSA when compared to that calculated with the theory outlined here.[27] However, when taking this correction into account, we compute a trapping volume for BSA of 0.441 µm$^3$ which, at a molar concentration of 500 pM, encloses only 0.133 BSA molecules. Once more, it seems unlikely that this minute number of molecules is able to produce the strong and efficient SERS response shown in the Fig. 4(a). Let us also not forget that the equations provided therein are valid for spherical analytes only. While similar expressions can be derived for non-spherical particles,[88] the typical shape of a nanoscale analyte usually differs from these ideal configurations and, for molecules, also exhibits a strong dependence on the experimental conditions. BSA, for example, can be considered a globular, i.e. spherical, protein with a hydrodynamic radius of 3.8 nm for a pH between 4.5 and 7, but begins to unfold at a more basic pH larger than 8.[27] All these effects are not taken into consideration by the current theory and represent a significant limitation for its predictive power. This clearly shows the unsatisfactory state of DEP nanoscale theory and we hope that these experimental results will stimulate fruitful theoretical investigations on the subject to guide future experimental efforts.

On the experimental side, we point out that the overall reproducibility of DEP biomolecular trapping was found to be lower than that for particle trapping. We took care to test at least 20 different devices to manipulate BSA and rhodamine separately and, while trapping of BSA was observed in about 85% of the tested devices, we were able to observe a signal from rhodamine in only 70% of cases. This reduced efficiency can be ascribed to several concurrent factors, the most important of which is surely the tinier size of rhodamine compared to BSA, and of these biomolecules when compared to a gold nanoparticles. A smaller size generally implies a smaller polarizability, resulting in a weaker trapping force. Furthermore, in the case of BSA, thermally- and electrically-induced flows might have also hampered the efficient delivery of the protein to the sensing hotspot. As for the former, it was hard to simulate the expected temperature increase in our device due to its extended size and high aspect ratio geometry. However, thanks to the low laser power employed during the SERS characterization, we expect that optical heating be significantly reduced in our device while, on the other hand, Joule heating might still play a role. In general, we can say that heating in our device is lower than that in standard plasmonic sensors – due to the presence of the electrical interconnections – but still higher than that provided by a continuous plasmonic film, and might indeed be a limiting factor in our experiments. As for any unwanted electrokinetic effect, we took care to minimize these by employing a low-conductivity buffer, as explained in the Methods section.

# Conclusions and outlook

To conclude, we have demonstrated that the simultaneous exploitation of low and high frequency electric fields provides a tremendous improvement in the trapping efficiency of plasmonic tweezers and in the sensing capabilities of DEP systems. This allowed us to manipulate molecules as small as Rhodamine B, with a mass of only 479 Da, thus extending the applications of tweezing devices to systems composed of less than one hundred atoms. This combination of plasmonic and DEP devices unlocks the manipulation and the study of virtually any known protein and oligopeptide,[63] thus providing a powerful platform to probe biological processes down to the single-molecule level, with far-reaching implications for all those domains at the crossroad between molecular biology, nanoscale medicine and analytical chemistry.

## Methods

The plasmonic electrodes are designed and fabricated following a strategy described in detail elsewhere,[61,62] where we show the benefits of employing an organic silane adhesion layer to counteract the effects of surface forces and to stabilize the electrodes.

The DEP trapping of 40 nm gold nanoparticles is performed by biasing the electrodes with a sinusoidal voltage having an amplitude of 7 V and a frequency of 1 MHz. Optical trapping is performed with a $\lambda$ = 974 nm laser, linearly polarized along the dimer axis, focused on the tweezer to a final power density of 3 mW/$\mu m^2$. The particles (from BBI Solutions) are diluted in filtered water just before use.

The DEP trapping of BSA and rhodamine is performed by applying a sinusoidal DEP trapping voltage with an amplitude of 7 V and a frequency of 300 kHz, while the Raman scattering is probed with a $\lambda$ = 974 nm laser focused on the tweezer to a final power density of 300 $\mu W/\mu m^2$. BSA molecules (from Merck) are diluted in a 5 mM HEPES buffer (from Merck) to different concentrations ranging from 10 $\mu M$ to 500 pM. The pH of the freshly made buffer, which usually lies between 4.5 and 5, is adjusted to a value between 7.5 and 8 with the addition of a 1 M solution of NaOH (from Merck). The buffer conductivity was measured to be between 100 and 500 $\mu S/cm$. As for rhodamine, Rhodamine B (from Merck) is simply diluted in water.

An Olympus IX73 inverted microscope coupled with an Andor Kymera 328i-A spectrograph equipped with a Newton 920 CCD is used to collect the experimental data, which are subsequently analyzed and plotted using Origin 2018.

Simulations are performed with COMSOL Multiphysics 5.6, with mesh element sizes ranging from 5 nm around the disks to 180 nm in the surrounding media.

## Acknowledgement

We gratefully acknowledge funding from the European Research Council (ERC-2015-AdG695206 Nanofactory). We are thankful to Dr. Siarhei Zavatski for his help with the HEPES buffer preparation.## References

# Electromagnetic manipulation of sub-500 Da biomolecules - Supporting Information

**Marco Riccardi** and Olivier J.F. Martin

*Laboratory of Nanophotonics and Metrology, Swiss Federal Institute of Technology,*

*CH-1015, Lausanne*

E-mail: marcori94@gmail.com

## Field maps

Figure S1 shows the low and high frequency field and intensity gradient distributions around the plasmonic dimer as calculated with COMSOL Multiphysics 5.6. Here and throughout the rest of this work, the DEP field, shown here on the left, is computed by biasing the disks with a sinusoidal voltage with amplitude of 7 V and frequency of 1 MHz. The optical field, shown here on the right, is calculated for an exciting *x*-polarized plane wave propagating along *z* with a wavelength λ = 974 nm and a power density of 3 mW/µm$^2$. It is evident that, under these conditions, DEP is able to generate a more intense field that extends further away from the electrodes and therefore results in the creation of stronger and longer-range trapping forces.

## Device fabrication

We realize a set of plasmonic electrodes by connecting a gold plasmonic disk dimer to a pair of gold nanorods that allow the electrical biasing of the disks. The nanorods are inserted inside the disks in such a way to avoid any electrical coupling between the two, which might affect the plasmonic resonance to be exploited for biosensing.[1,2] The whole structure is then fabricated following a dedicated electron-beam liftoff process that allows the creation of the extremely high aspect ratio electrical connectors required by this device, without incurring into any peeling of the structures.[3] This is achieved by employing MPTMS as an organic adhesion layer between the silica substrate and the gold structures, therefore providing a stronger adhesion force between these two materials that ensures long-term stability of the device.[3] The fabricated structures are shown in Fig. S2(a), which confirms the realization of several plasmonic electrodes arranged into an interdigitated electrode array geometry.

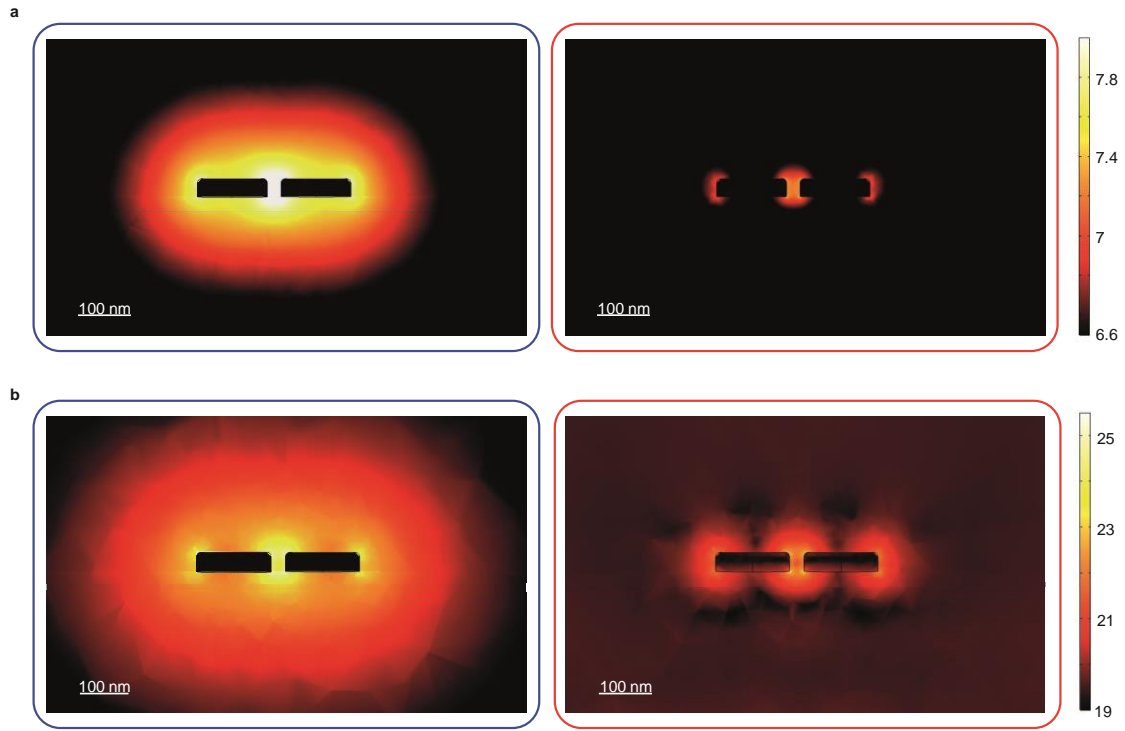

**Fig. S1:** Map of the electric field (a) and electric field intensity gradient (b) for DEP (left) and optical (right) field around the structures. The color legends represent the logarithm of the plotted quantity.

The final gold cylinders have a diameter of about 140 nm and a height of 40 nm, while the gap between two adjacent cylinders is 30 nm. Figure S2(b) shows the results of the electrical and optical characterizations of one device, which demonstrate an excellent electrical insulation between the electrodes and the presence of a plasmonic resonance at around λ = 1000 nm, in agreement with the simulated behaviour.

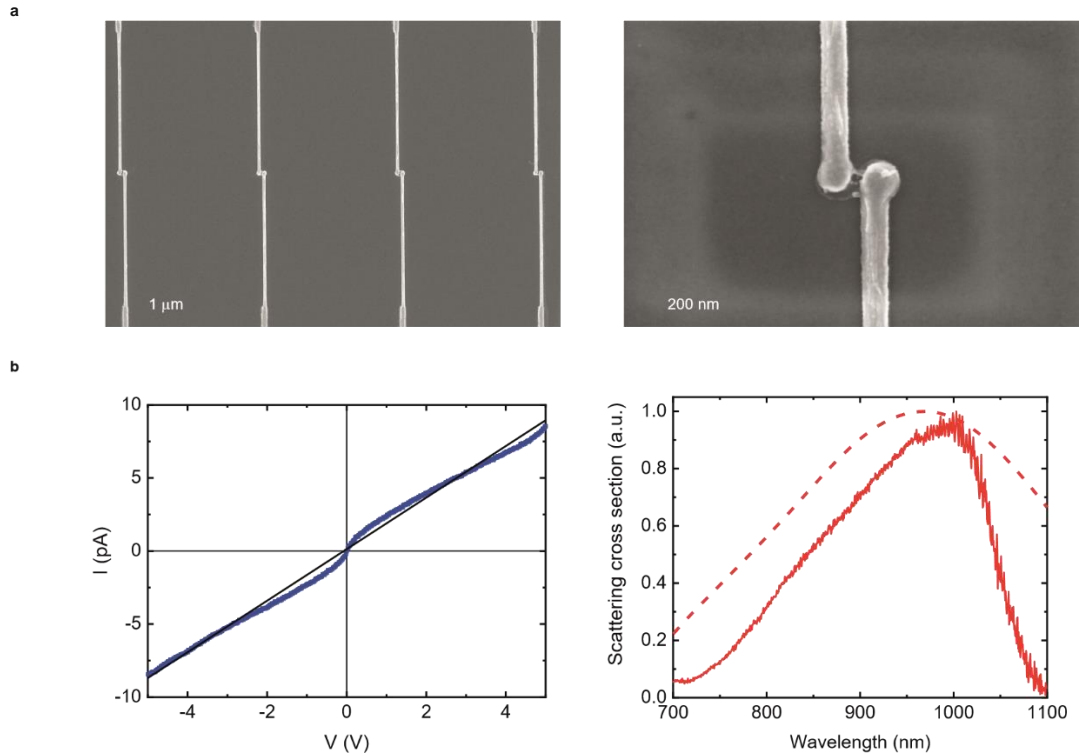

**Fig. S2:** (a) SEM images of the fabricated structures showing, on the left, the interdigitated plasmonic electrode array. On the right, a zoomed-in view of a typical structure is shown. Each disk has a diameter of 150 nm and the gap between two adjacent disks is 30 nm. (b) IV curve (left) and scattering cross section (right) of one typical device. The latter are measured in a water environment, with the dashed line representing the theoretical cross section calculated with COMSOL Multiphysics 5.6.

## Refractive index sensing of BSA

Figure S3 demonstrates the redshift of the plasmonic resonance of the electrodes upon DEP trapping (7 V, 300 kHz) of BSA proteins. These are diluted in a HEPES buffer to a concentration of 10 µM. We see here that, as soon as the DEP field is turned on, the resonance clearly moves towards lower energies, confirming the trapping of BSA molecules. These induce a less abrupt shift than the nanoparticles due to their smaller size and we therefore witness a more gradual and constant shift, as more and more molecules are trapped in the vicinity of the disks, rather than the step-like behaviour shown in the main document for nanoparticles.

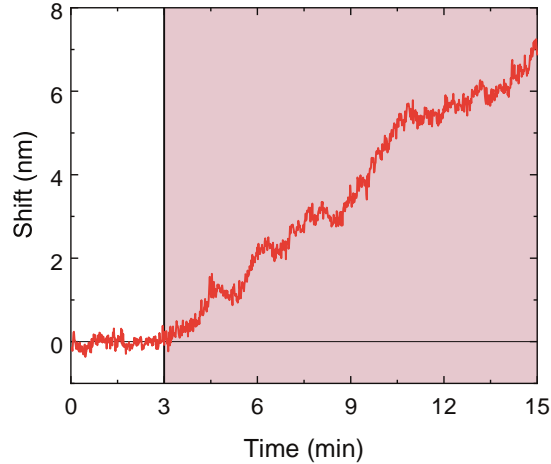

**Fig. S3:** Evolution of the position of the plasmonic resonance of the cylinders over time. Upon the activation of DEP (pink shaded areas), the resonance gradually moves toward longer wavelengths.

# Determination of molecular diffusion coefficients from SERS intensity data

The time evolution of the molecular concentration $\phi$ of an analyte diffusing away from a substrate at $z = 0$ is given by[4]

$$\phi(z=0, t>0) = \frac{2}{\sqrt{4\pi D t}}, \tag{1}$$

where $D$ is the molecular diffusion coefficient. Since $\phi$ is proportional to the SERS intensity, Eq. (1) can be used to fit the SERS data recorded after the DEP trapping is turned off. To this end, Fig. S4 presents the fits of the data shown in Fig. 4 of the main document, and allows the extraction of the diffusion coefficients of both BSA and rhodamine. These are found to be about $6 \cdot 10^{-7}$ m²/s for the former and $8 \cdot 10^{-10}$ m²/s for the latter.

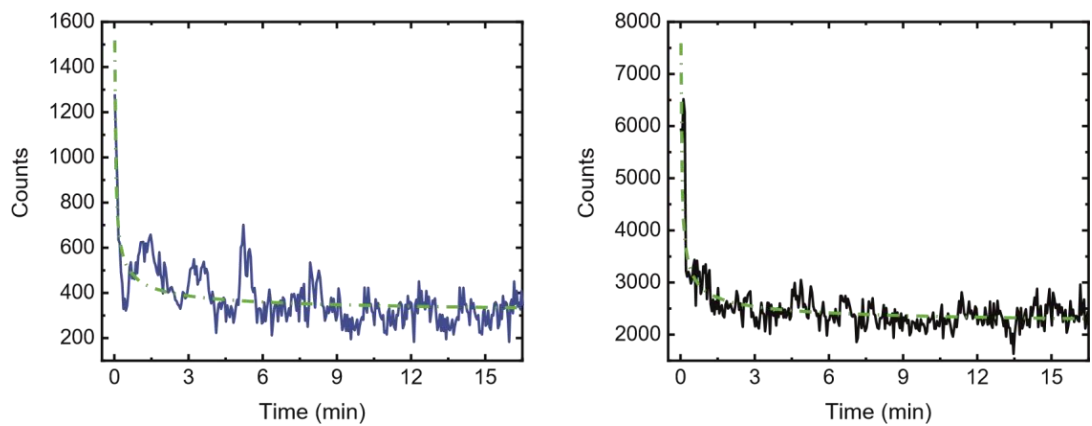

**Fig. S4:** Best fitting curve (in green), according to Eq. (1), for both BSA (left) and Rhodamine (right).

## Empirical determination of the particles trapping volume

We present in Fig. S5 a series of snapshots, from the video provided with the manuscript, that shows the DEP trapping of a 40 nm gold nanoparticle when the electrodes are biased with a 1 MHz voltage having an amplitude of 7 V. With reference to the scale bar and the electrode highlighted by the yellow circle, it can be clearly appreciated how the DEP field is able to trap a gold nanoparticle even when this is about 3 µm away from the disk dimer, i.e. the bright spot in the middle of the circle. This results in a trapping volume of 56.5 µm$^3$, more than 160 times greater than that depicted in Fig. 2d in the main text and calculated with the theory described in the main text. Let us also point out that this is an empirical value that should be considered as the *minimum* size of the trapping volume, while the real one is likely to be much larger. We attribute this mismatch between the predicted and observed volumes to the presence of a stabilizing negative charge on the surface of the particle, which generates a polarized electric double layer at the particle-water interface. This, in turn, contributes to an increase of the total polarization of the particle and therefore enlarges its trapping volume. This additional surface contribution to the total dipole moment, which is of the utmost importance in nanoscale analytes that have an increased surface to volume ratio, is not taken into account by the theory described in the main text and therefore makes it unsuited to describe the induced polarization of nanoscale objects.

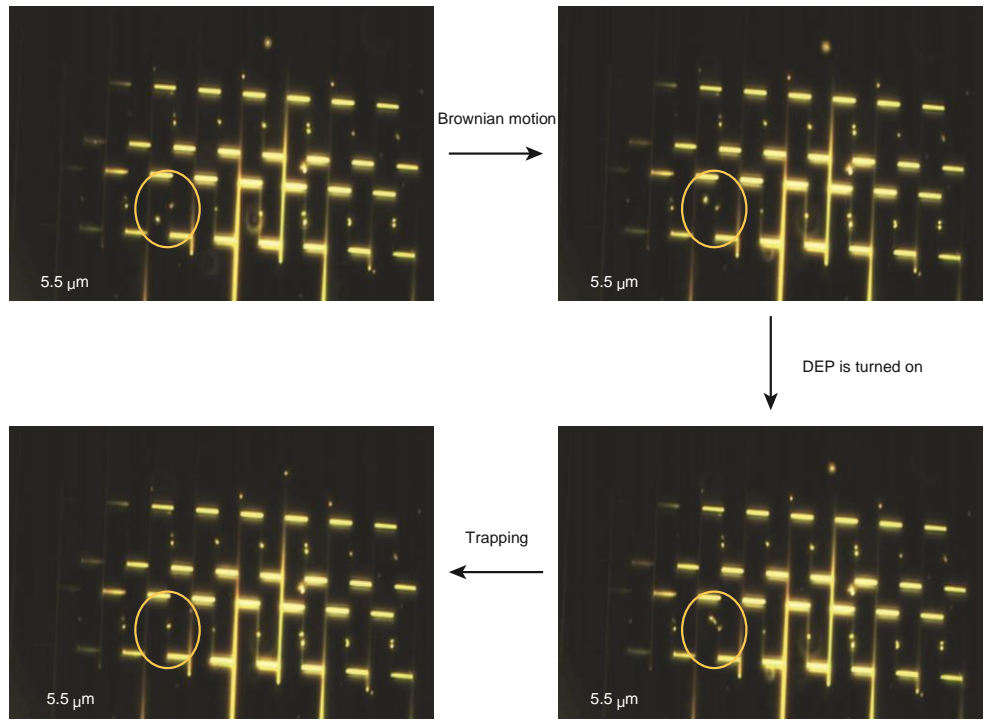

**Fig. S5:** Dark field optical images of a particle trapping event. The DEP field readily traps the particle in the yellow circle even when this is several microns away from the electrodes. The electrode design is an older version of that presented in Fig. S2.